# On the connection between angular and partial waves time delays in electron scattering

M. Ya. Amusia[1,2], A. S. Baltenkov[3]
and
I. Woiciechowski[4]

[1] Racah Institute of Physics, the Hebrew University, Jerusalem 91904, Israel
[2] Ioffe Physical-Technical Institute, St. Petersburg 194021, Russia
[3] Arifov Institute of Ion-Plasma and Laser Technologies,
100125, Tashkent, Uzbekistan
[4] Alderson Broaddus University, WV 26416, Philippi, USA

**Abstract**

Within the framework of a Dirac bubble potential model for the $C_{60}$ fullerene shell we investigated the *angular time delay* in slow-electron elastic scattering by $C_{60}$ as well as *average time delay* of electrons in this process. It is demonstrated how the angular time delay is connected to the Eisenbud-Wigner-Smith (EWS) time delay. The angular and energy dependences of these times are investigated. The studies conducted shed light to some extent on the specific features of these dependencies.

## 1. Introduction.

The domain of applicability of the so-called *angular time delay*, introduced by Froissart, Goldberger and Watson in paper [1] is considerable broader than the *partial-wave time delay* introduced by Eisenbud [2], Wigner [3] and Smith [4]. In [1] by the direct generalization of the Eisenbud-Wigner-Smith (EWS) partial wave time delay for scattering of the wave packet in pure eigenstates of the $S$ matrix, the following expression for the angular time delay of the packet scattered by a spherical target in the direction $\vartheta$ was found

$$\Delta t(k,\vartheta) = \hbar \frac{\partial}{\partial E} \arg f(k,\vartheta), \quad \vartheta \neq 0. \quad (1)$$

Here $f(k,\vartheta)$ denotes the total scattering amplitude. According to (1) the forward scattering $\vartheta = 0$ must be excluded. This is due to the inevitable interference effects between the forward scattered wave and the incident wave that gives rise to the optical theorem [5]. Equation (1) serves as the basis for describing the temporal picture of atomic and molecular photoionization processes [6-12]. The above equation in the case of scattering needs to be modified to exclude $\vartheta \neq 0$ while the problem of interference with the non-scattered wave does not arise in the context of some other processes, e.g. photoionization.

Note that the usual experimental set-up of electron collisions permits measurements of angular time delay but not the EWS times.

The rapidly oscillatory contributions in angular dependence of $\Delta t(k,\vartheta)$, with the frequency of the order of the inverse free time of flight across the de Broglie wavelength in the function (1) are related to the uncertainty principle. According to Nussenzveig [13-15], therefore it is necessary to get rid of them by averaging the function $\Delta t(k,\vartheta)$ over a distance in angle that is of the order of the de Broglie wavelength. The result is the *average time delay in the scattering process* $\langle \Delta t(k) \rangle$ that is readily interpretable in terms of the average time of flight for a classical free particle beam. The general formulas describing the average time delay are derived in [14] without introduction into consideration the usual partial-wave expansion of the scattering amplitude that permits to characterize the scattering process by a set of phases $\delta_l$ that are function of the incoming particle linear momentum $k$:

$$f(k,\vartheta) = \frac{1}{2ik}\sum_l (2l+1)(e^{2i\delta_l} - 1)P_l(\cos\vartheta). \tag{2}$$

Therefore, the explicit form of $\Delta t(k,\vartheta)$ in (1) as a function of the scattering angle $\vartheta$ of the plane-wave packet and the electron momentum $k$ remains unknown until they are not expressed via $\delta_l$. The aim of this paper is, while remaining within the framework of the Dirac bubble potential model for the $C_{60}$ fullerene shell [16] and using the scattering phase shifts $\delta_l(k)$ and EWS partial wave times delay,

$$\tau_l(k) = 2\hbar\frac{\partial \delta_l}{\partial E}, \tag{3}$$

obtained in this paper, to calculate in the explicit form the function $\Delta t(k,\vartheta)$ for slow-electron elastic scattering by $C_{60}$. The attractiveness of the bubble potential model is in its ability to present analytically results on phase shifts and cross sections.

In the next Section the $\vartheta$-dependence of the angular time delay $\Delta t(k,\vartheta)$ for some fixed electron momentum $k$ is investigated. In Section 2 this function is studied as a function of $k$ for some fixed polar scattering angles $\vartheta$ of the incident plane wave train. In the last Section 3, the function $\Delta t(k,\vartheta)$ is averaged over a distance of the order of the de Broglie wavelength and the *average angular time delay* $\langle \Delta t(k) \rangle$ is obtained and expressed via the partial wave time-delay.

2.  **Angular dependence of function $\Delta t(k,\vartheta)$.**

The argument of scattering amplitude $f(k,\vartheta)$ is determined by the ratio of the imaginary part of function (2), $\text{Im} f(k,\vartheta)$, to its real part, $\text{Re} f(k,\vartheta)$ as

$$\arg f(k,\vartheta) = \arctan\frac{\text{Im} f(k,\vartheta)}{\text{Re} f(k,\vartheta)}, \tag{4}$$

whereas angular time delay (1) is described by the following general expression [17]

$$\Delta t(k,\vartheta) = \frac{(\text{Im} f)'(\text{Re} f) - (\text{Re} f)'(\text{Im} f)}{|f|^2}. \tag{5}$$

From here on we use the atomic system of units. The prime in (5) and further in other formulas denote differentiation with respect to the kinetic energy of an electron $E$. Let us first consider the case when all phase shifts in (1), with the exception of $\delta_l(k)$, are equal to zero. In this case

$$\Delta t(k,\vartheta) = \delta'_l, \tag{6}$$

that is, the angular time delay is isotropic, being independent upon the angle $\vartheta$, and is equal to half of the EWS partial wave time delay (3).

Let us make a next step and assume that in the electron scattering amplitude (2) only two scattering phases $\delta_l$ and $\delta_{l'}$ are nonzero. In this case, the angular time delay (5) is determined by the following combination of the Legendre polynomials

$$\Delta t(k,\vartheta)_{ll'} = \frac{\sum_{i=l,l'}\sum_{j=l,l'}[(2i+1)(2j+1)P_i P_j \sin(2\delta_i - \delta_j)\sin\delta_j]\delta_i'}{\sum_{i=l,l'}\sum_{j=l,l'}(2i+1)(2j+1)P_i P_j \sin\delta_i \sin\delta_j \cos(\delta_i - \delta_j)}. \quad (7)$$

Figure 1($P_0,P_1$) shows the angular time delay $\Delta t(k,\vartheta)_{01}$, calculated with phases $\delta_l \equiv \delta_l(k)$ and their derivatives $\delta' \equiv \delta_l'(k)$ from article [16], that depends upon Legendre polynomials $P_0(\cos\vartheta)$ and $P_1(\cos\vartheta)$

$$\Delta t(k,\vartheta)_{01} =$$
$$\frac{P_0[P_0 \sin^2\delta_0 + 3P_1 \sin(2\delta_0 - \delta_1)\sin\delta_1]\delta_0' + 3P_1[3P_1 \sin^2\delta_1 + P_0 \sin(2\delta_1 - \delta_0)\sin\delta_0]\delta_1'}{P_0^2 \sin^2\delta_0 + 6P_0 P_1 \sin\delta_0 \sin\delta_1 \cos(\delta_0 - \delta_1) + 9P_1^2 \sin^2\delta_1} \quad (8)$$

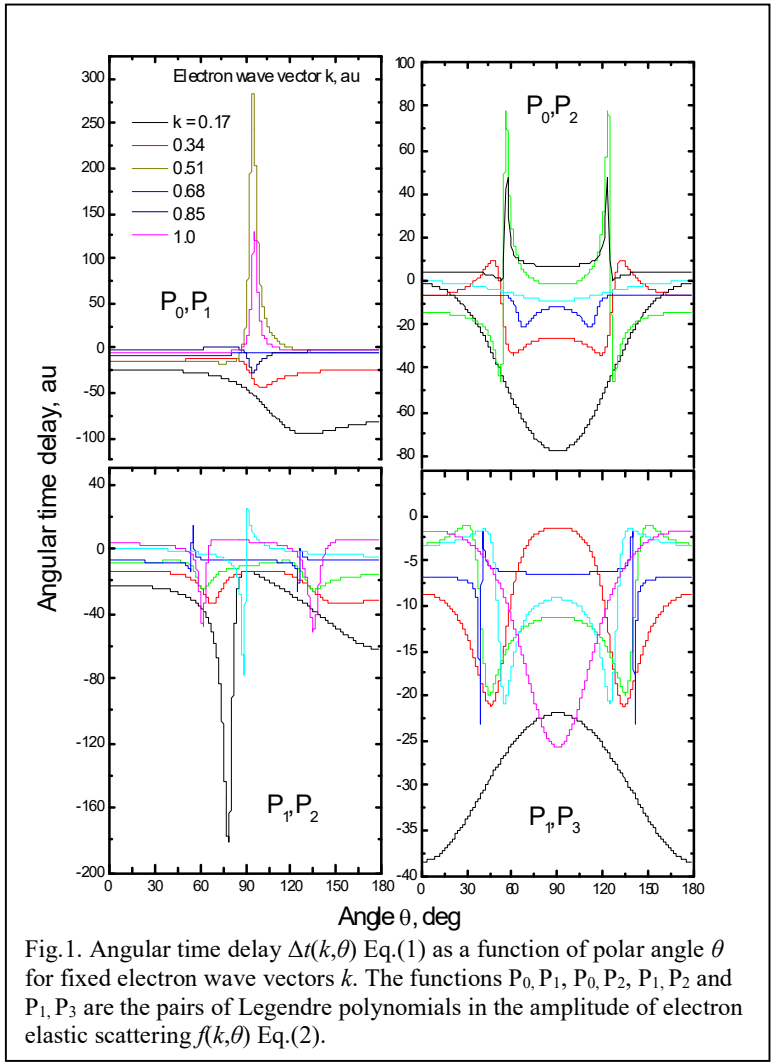

Fig.1. Angular time delay $\Delta t(k,\theta)$ Eq.(1) as a function of polar angle $\theta$ for fixed electron wave vectors $k$. The functions $P_0, P_1$, $P_0, P_2$, $P_1, P_2$ and $P_1, P_3$ are the pairs of Legendre polynomials in the amplitude of electron elastic scattering $f(k,\theta)$ Eq.(2).

as a function of the scattering angle $\vartheta$ at some fixed values of electron linear momenta $k = \sqrt{2E}$. Angular time delays in these figures are given in the atomic units equal to 24.2 as, where 1 as = $10^{-18}$ s.

According to figure 1($P_0,P_1$) at low electron moments $k = 0.17$ and 0.34, the time delay of the scattering packet is negative at all scattering angles and close to zero at $k = 0.68$ and 0.85. At momenta $k = 0.51$ and $k = 1.0$, the time delays reach a maximum, $\approx 298$ atomic units (au) at $\vartheta = 95°$ in the first case and $\approx 140$ au at the same angle in the second one. The appearance of these sharp peaks in the curves in Fig.1($P_0,P_1$) is due to the almost vanishing of the denominator in Eq. (8).

Let us now assume again that in the electron scattering amplitude (2) only two scattering phases are nonzero, $\delta_0(k)$ and $\delta_2(k)$, with $\delta_2(k)$ substituting $\delta_1(k)$. In this case, the angular time delay (5) is determined by the combination of the Legendre polynomials $P_0(\cos\vartheta)$ and $P_2(\cos\vartheta)$.

$$\Delta t(k,\vartheta)_{02} = \frac{P_0[P_0 \sin^2\delta_0 + 5P_2 \sin(2\delta_0 - \delta_2)\sin\delta_2]\delta_0' + 5P_2[5P_2 \sin^2\delta_2 + P_0 \sin(2\delta_2 - \delta_0)\sin\delta_0]\delta_2'}{P_0^2 \sin^2\delta_0 + 10P_0P_2 \sin\delta_0 \sin\delta_2 \cos(\delta_0 - \delta_2) + 25P_2^2 \sin^2\delta_2}. \quad (9)$$

Since the sum of the orbital moments (indices of the Legendre polynomials) is an even number, the curves $\Delta t(k,\vartheta)_{02}$ in Fig. 1(P$_0$,P$_2$) are symmetric functions relative to the angle $\vartheta = 90°$. The 3D-picture of the function $\Delta t(k,\vartheta)_{01}$ is a figure of rotation of this curve around the polar axis, along which an incident plane wave train hits the target. The "wings of the star" shown there correspond to polar scattering angles $\vartheta = 57°$ and $123°$.

Fig. 1(P$_1$,P$_2$) depicts the curves corresponding to the pair of polynomials $P_1(\cos\vartheta)$ and $P_2(\cos\vartheta)$. Since the sum of their indexes is an odd number, the set of curves in this figure is qualitatively similar to the set of curves in Fig. 1(P$_0$,P$_1$). However, due to the increase in the degree of the cosine of the scattering angle $\vartheta$ in the formula

$$\Delta t(k,\vartheta)_{12} = \frac{3P_1[3P_1 \sin^2\delta_1 + 5P_2 \sin(2\delta_1 - \delta_2)\sin\delta_2]\delta_1' + 5P_2[5P_2 \sin^2\delta_2 + 3P_1 \sin(2\delta_2 - \delta_1)\sin\delta_1]\delta_2'}{9P_1^2 \sin^2\delta_1 + 30P_1P_2 \sin\delta_1 \sin\delta_2 \cos(\delta_1 - \delta_2) + 25P_2^2 \sin^2\delta_2}, \quad (10)$$

the frequency of the curves oscillations is significantly higher than that of the curves in Fig. 1(P$_0$,P$_1$).

And, finally, completing the study of the angular dependence of the time delay $\Delta t(k,\vartheta)$ for the case when only a pair of phase shifts is different from zero, we present in Fig. 1(P$_1$,P$_3$) the following combination of the Legendre polynomials $P_1(\cos\vartheta)$ and $P_3(\cos\vartheta)$

$$\Delta t(k,\vartheta)_{13} = \frac{3P_1[3P_1 \sin^2\delta_1 + 7P_3 \sin(2\delta_1 - \delta_3)\sin\delta_3]\delta_1' + 7P_3[7P_3 \sin^2\delta_3 + 3P_1 \sin(2\delta_3 - \delta_1)\sin\delta_1]\delta_3'}{9P_1^2 \sin^2\delta_1 + 42P_1P_3 \sin\delta_1 \sin\delta_3 \cos(\delta_1 - \delta_3) + 49P_3^2 \sin^2\delta_3}, \quad (11)$$

The set of curves in this figure is qualitatively similar to that in Fig. 1(P$_0$,P$_2$). But in this case, all curves, including the curve k = 1.0, lie in the negative half-plane.

According to Fig. 1, the angular dependences of the function $\Delta t(k,\vartheta)$ are nontrivial rapidly oscillating curves lying at low electron energies, mainly in the negative half-plane. The situation changes with increasing electron energy.

3. ***k*-dependence of angular time delay** $\Delta t(k,\vartheta)$.

Let us investigate the angular time delay as a function of electron linear momentum $k$ for some fixed values of polar angles $\vartheta$. The calculation results are presenter in Fig. 2. The result obtained using by formula (8) are shown in Fig. 2(P$_0$,P$_1$). All curves in this figure at small electron momenta tend to infinity. The reason for this is that the scattering phase shift in a short-range Dirac-bubble potential should follow the Wigner threshold law $\delta_l(E) \propto E^{l+1/2}$ [18]. In the case of *s*-phase shift we have $\delta_0(E \to 0) \propto \pi - E^{1/2}$.

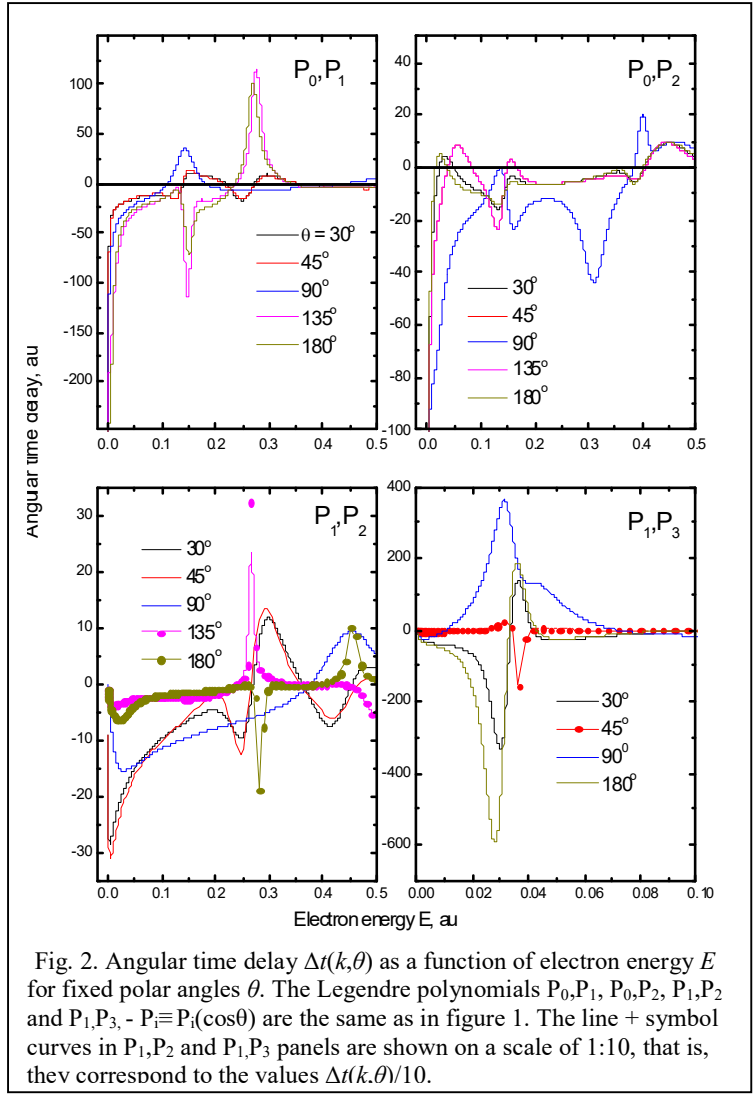

Fig. 2. Angular time delay $\Delta t(k,\theta)$ as a function of electron energy $E$ for fixed polar angles $\theta$. The Legendre polynomials $P_0,P_1$, $P_0,P_2$, $P_1,P_2$ and $P_1,P_3$, - $P_i \equiv P_i(\cos\theta)$ are the same as in figure 1. The line + symbol curves in $P_1,P_2$ and $P_1,P_3$ panels are shown on a scale of 1:10, that is, they correspond to the values $\Delta t(k,\theta)/10$.

So, the angular time delay $\Delta t(k,\vartheta)_{01}$, that contains derivative of s-phase shift, for $k\to 0$ goes to infinity since $\delta'_0(E\to 0) \propto -E^{-1/2}$. For the orbital moments $l>0$ the phase shifts derivative vanishes at zero energy as $\delta'_l(E\to 0) \propto E^{l-1/2}$. At small scattering angles $\vartheta$, up to an angle of 45°, the curves practically coincide with each other. Curves corresponding to angles 135° and 180° have alternating signs and are characterized by peaks in both positive and negative half-planes of the figure.

We see a qualitatively similar picture in Fig. 2($P_0,P_2$), where the curves for $\Delta t(k,\vartheta)_{02}$ are presented. Here also, the presence of the derivative of s-phase shift in formula (9) leads to infinity of this function at small electron momenta $k$. The curves for angles 30° and 180° almost coincide in this figure. The curve for $\vartheta = 90°$, among other curves, is characterized by the maximum amplitude of oscillations.

We observe a different threshold behavior of the curves in Fig. 2($P_1,P_2$) and Fig. 2($P_1,P_3$), which depict the functions $\Delta t(k,\vartheta)_{12}$ and $\Delta t(k,\vartheta)_{13}$. All curves vanish at $k\to 0$. In Fig. 2($P_1,P_2$), curves for 135° and 180° are shown on a scale of 1:10, since their maximum values reach several hundreds of atomic units. The amplitudes of the curves in Fig. 2($P_1,P_3$) reach about 600 au; curve 45° is shown on a scale of 1:10 also.

We limit ourselves to concrete examples of two non-zero phases in all the expansion of the scattered electron wave function (2) into partial waves. We see that this leads to a very complex, rapidly oscillating upon energy $E$ and scattering angle $\vartheta$ dependence. The increase of the number of scattering phases taken into account significantly affects the angular time delay picture making it so rapidly oscillating that averaging it over incoming electron energy and scattered angle becomes inevitable to make time delay observable in experiment.

4. **The average time delay in the scattering process.**

The average angular time delay is obtained from (1) by averaging over the energy spectrum of the incident wave packet, as well as over directions weighted by the differential cross section. As a result of such averaging, Nussenzweig [13] obtained the following expression

$$\frac{1}{\sigma_t(k)}\left(\int |f(k,\vartheta)|^2 \Delta t(k,\vartheta)d\Omega + \frac{2\pi}{k^2}\frac{d}{dE}[k\,\mathrm{Re}\,f(k,0)]\right) = \frac{\pi}{\sigma_t(k)k^2}\sum_l (2l+1)2\delta'_l = \langle \Delta t(k)\rangle. \quad (12)$$

Thus, the average time delay for the plane wave train $\langle \Delta t(k)\rangle$ is the sum of EWS partial waves times delays $\tau_l(k)$ (3). The second term on the left side of the equation (12) removes the contribution to the average angular time delay of the forward scattering.

The result of calculating the function $\langle \Delta t(k)\rangle$ is shown in Fig. 3. The same figure shows the curve calculated in article [16] under the assumption that the statistical weight of partial-wave components in the sum (12) is not equal to $\pi(2l+1)/\sigma_t k^2$, but is a ratio of the electron elastic scattering partial cross section upon $C_{60}$ and the total one. The time $\langle \Delta t(k)\rangle$, as in the case of the cross section for elastic scattering of electrons by $C_{60}$ [19], reaches a positive maximum within the range of the $f$-partial wave resonance.

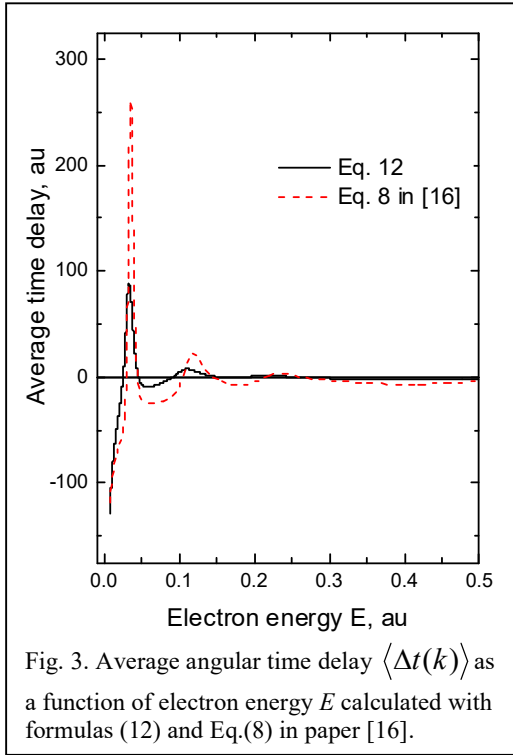

Fig. 3. Average angular time delay $\langle \Delta t(k)\rangle$ as a function of electron energy $E$ calculated with formulas (12) and Eq.(8) in paper [16].

## 5. Concluding remarks.

Using concrete example of electron scattering by fullerene $C_{60}$, we for the first time explicitly expressed the angular time-delay $\Delta t(k,\vartheta)$ via scattering phase shifts and their energy derivatives. We demonstrate the complexity of $\Delta t(k,\vartheta)$ as a function of incoming electron energy E and its scattering angle $\vartheta$. We see that $\Delta t(k,\vartheta)$ and even averaged over proper intervals of E and $\vartheta$ $\langle \Delta t(k)\rangle$ are more sensitive to the scattering phases than the absolute and even differential in angle scattering cross section since they depend not only on the cross section phases, but upon their derivatives in energy. This makes theoretical and experimental investigation of time delay a promising direction of research in the area of atomic scattering.

**Acknowledgment**
A. S. Baltenkov is grateful for the support to the Uzbek Foundation Award OT-2-46.

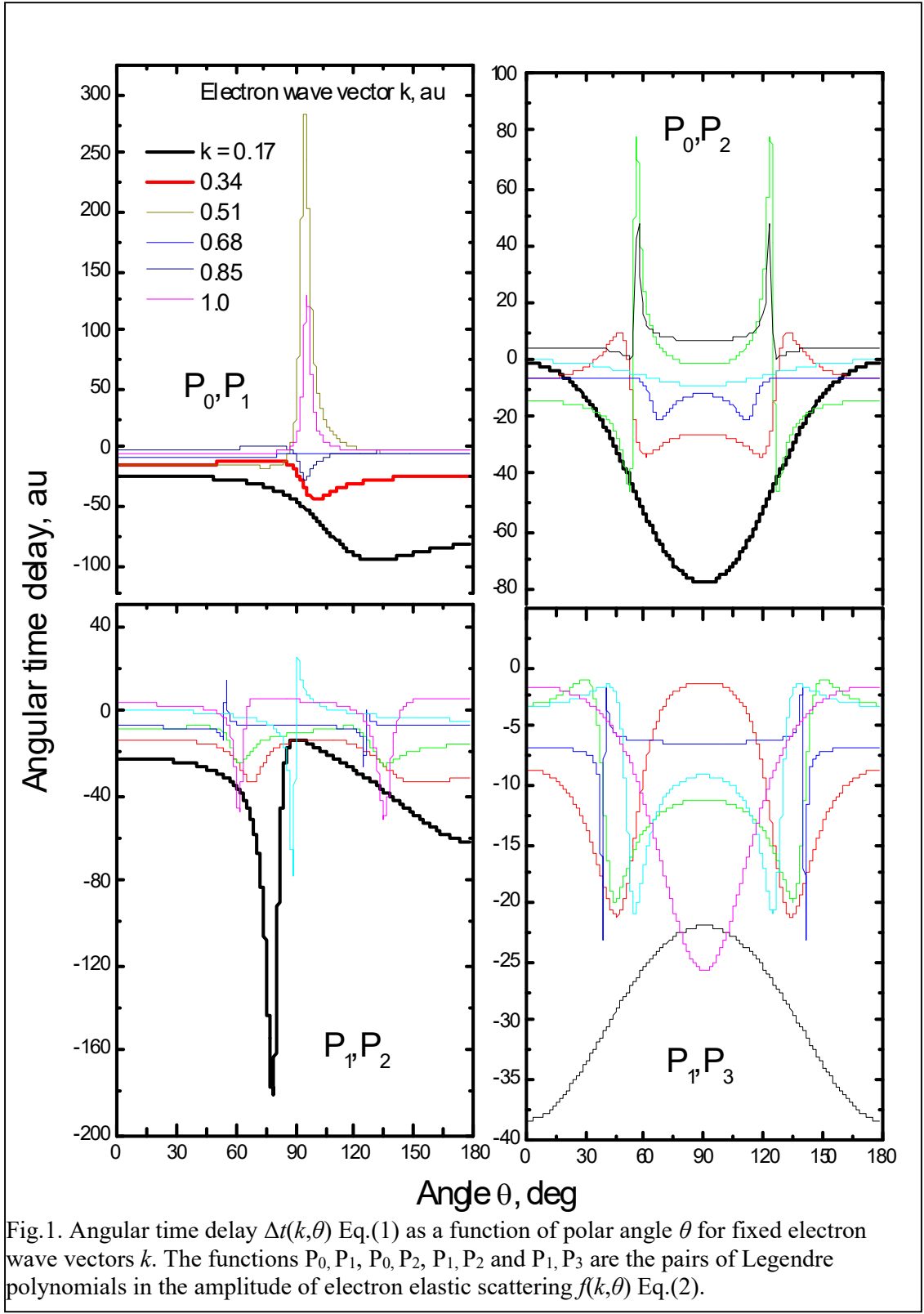

Fig.1. Angular time delay $\Delta t(k,\theta)$ Eq.(1) as a function of polar angle $\theta$ for fixed electron wave vectors $k$. The functions $P_0, P_1$, $P_0, P_2$, $P_1, P_2$ and $P_1, P_3$ are the pairs of Legendre polynomials in the amplitude of electron elastic scattering $f(k,\theta)$ Eq.(2).

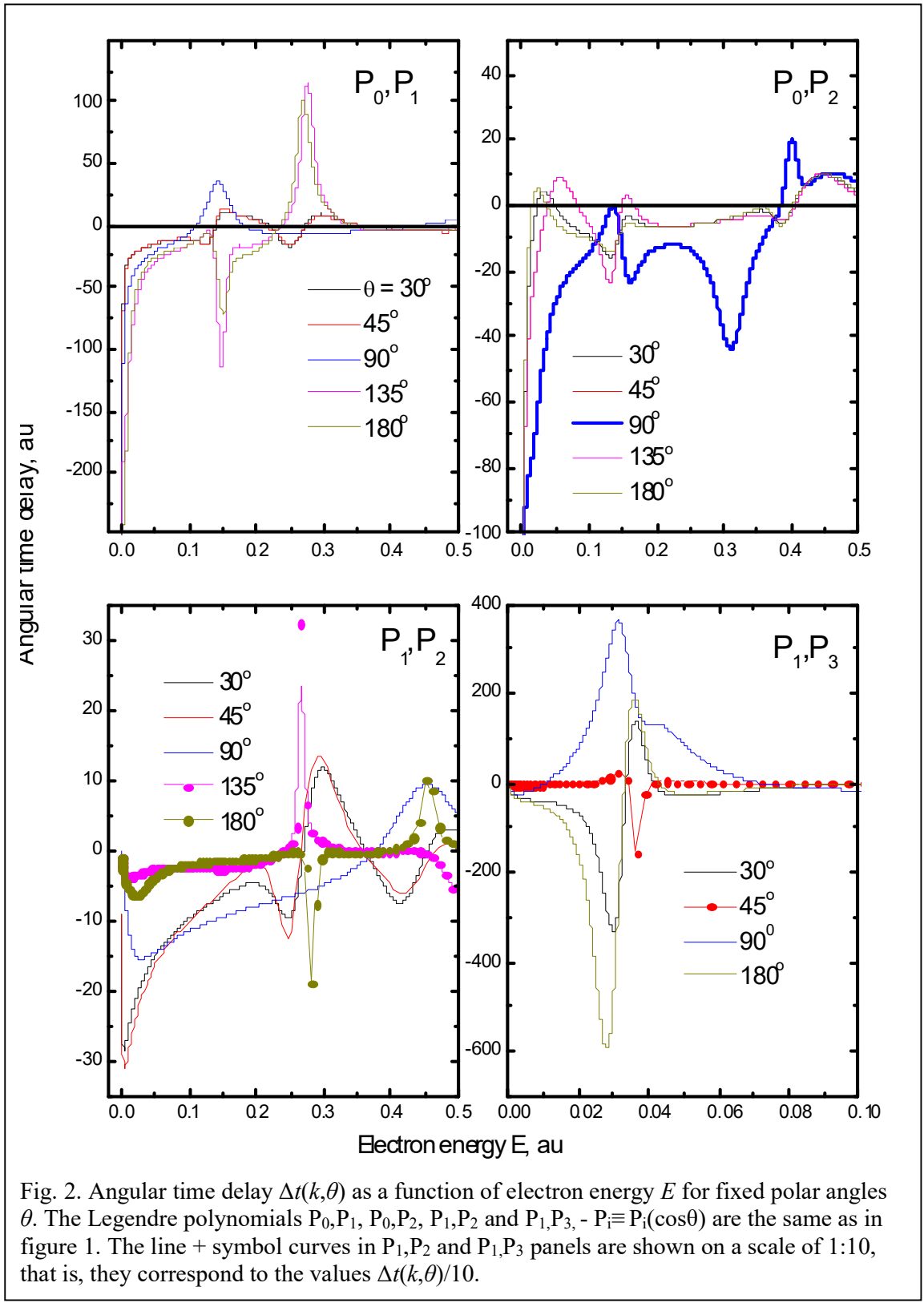

Fig. 2. Angular time delay $\Delta t(k,\theta)$ as a function of electron energy $E$ for fixed polar angles $\theta$. The Legendre polynomials $P_0,P_1$, $P_0,P_2$, $P_1,P_2$ and $P_1,P_3$, - $P_i \equiv P_i(\cos\theta)$ are the same as in figure 1. The line + symbol curves in $P_1,P_2$ and $P_1,P_3$ panels are shown on a scale of 1:10, that is, they correspond to the values $\Delta t(k,\theta)/10$.

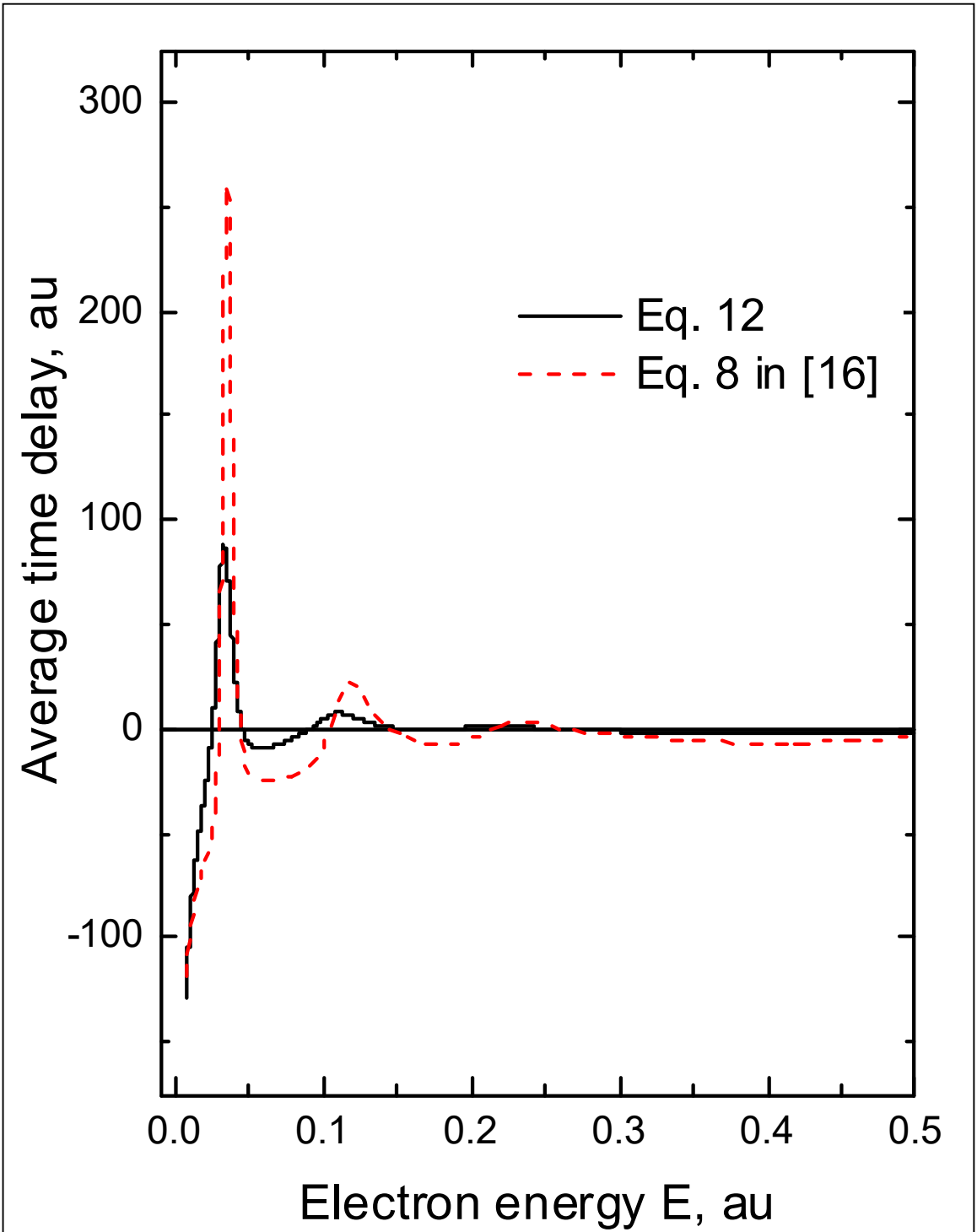

Fig. 3. Average angular time delay $\langle \Delta t(k) \rangle$ as a function of electron energy $E$ calculated with formulas (12) and Eq.(8) in paper [16].